\documentclass[english,aps,prl,twocolumn,amsmath,amssymb,showpacs,superscriptaddress,notitlepage,longbibliography]{revtex4-2}
\usepackage{bm}
\usepackage[pdftex]{graphicx,hyperref}
\hypersetup{colorlinks = true, urlcolor = blue, linkcolor = blue, citecolor = blue}
\usepackage{mathtools}
\usepackage{ulem}
\usepackage{color}

\begin{document}

\title{Altermagnetic Even-Odd Effects in CsV$_2$Te$_2$O Josephson Junctions}

\author{Chuang Li}
\thanks{These authors contributed equally.}
\affiliation{Center for Correlated Matter and School of Physics, Zhejiang University, Hangzhou 310058, China}

\author{Jin-Xing Hou}
\thanks{These authors contributed equally.}
\affiliation{Hefei National Laboratory, Hefei, Anhui, 230088, China}
\affiliation{International Center for Quantum Design of Functional Materials (ICQD), University of Science and Technology of China, Hefei, Anhui 230026, China}

\author{Shuai-Ling Zhu}
\affiliation{Center for Correlated Matter and School of Physics, Zhejiang University, Hangzhou 310058, China}

\author{Hao Zheng}
\affiliation{Center for Correlated Matter and School of Physics, Zhejiang University, Hangzhou 310058, China}

\author{Yu Song}
\affiliation{Center for Correlated Matter and School of Physics, Zhejiang University, Hangzhou 310058, China}

\author{Yang Liu}
\affiliation{Center for Correlated Matter and School of Physics, Zhejiang University, Hangzhou 310058, China}

\author{Song-Bo Zhang}
\email{songbozhang@ustc.edu.cn}
\affiliation{Hefei National Laboratory, Hefei, Anhui, 230088, China}
\affiliation{International Center for Quantum Design of Functional Materials (ICQD), University of Science and Technology of China, Hefei, Anhui 230026, China}

\author{Lun-Hui Hu}
\email{lunhui@zju.edu.cn}
\affiliation{Center for Correlated Matter and School of Physics, Zhejiang University, Hangzhou 310058, China}

\begin{abstract}
The interplay between conventional superconductivity and unconventional magnetism offers an exciting platform for realizing exotic superconducting phenomena. Here, we investigate Josephson effects in planar and vertical junctions based on CsV$_2$Te$_2$O-family materials, which host hidden $d$-wave altermagnetism with G‑type antiferromagnetic order. In monolayer-based planar junctions, the quasi-1D, nearly flat, spin-polarized bands of the altermagnet, when coupled to $s$-wave superconductors, produce a \textit{fully} spin-polarized supercurrent with strong directional anisotropy—a spin-selective Josephson effect. In multilayers, we uncover an \textit{altermagnetic even-odd effect}: spin-polarized supercurrents persist only in odd-layer planar junctions but cancel exactly in even layers. Thus, layer parity acts as a switch for spin-polarized supercurrent. In vertical junctions, odd-layer barriers enhance equal-spin triplet transport while even layers favor opposite-spin transport, yielding a robust period-two oscillation in the total supercurrent with layer number. These layer‑parity-dependent responses represent a general even‑odd effect in hidden altermagnets, applicable to diverse magnetic and transport phenomena.
\end{abstract}

\maketitle

\noindent{\bf Introduction}\\
Unconventional magnetism has emerged as a cornerstone of condensed matter physics~\cite{Hirsch1990prb,Ikeda1998prl,wu2004prl,CJWu07PRB,Chen2014Anomalous,nakatsuji2015large,noda2016momentum,Ahn2019prb,liu2025different,liu2025prx,liu2025symmetry}. A pivotal development in this field is the recent identification of collinear altermagnetism~\cite{vsmejkal2020crystal,Naka19NC,Nakaprb,hayami2019momentum,Yuan2020Giant,mazin2021prediction,ma2021multifunctional,shao2021spin,ifmmode2022Beyond,ifmmode2022Emerging}, which has since spurred intense theoretical interest~\cite{Ouassou2023prl,Zhang2024NatComm,Beenakker2023prb,Cheng2024prb,maierPRB2023weak,he2023prl,du2023topological,leeb2024prl,yuan2024prl,Roig2024prb,liu2024twist,gu2025ferroelectric,Xu2025altermagnetic,Chakraborty2025prl,duan2025prl,Lin2025prl,wang2025prl,vila2024orbital,Sim2025pair,wu2025intra,ZhuXC25PRL,ChenR25PRL,Zhang2025xtype} and triggered numerous experimental investigations~\cite{Feng2022natele,gonzalez2023prl,lee2024MnTe,krempasky2024altermagnetic,osumi2024MnTe,lu2025signature,fedchenko2024ruo2,lin2024ruo2,liu2024chiral,reimers2024CrSb,ding2024CrSb,han2024electrical,zhou2025manipulation,yang2025three,Regmi2025natcomm,zhou2026surfaces,shao2026epita}. Unlike conventional N\'eel antiferromagnets, altermagnets are characterized by a compensated magnetic order that hosts momentum-dependent spin polarization, opening a distinct route to spin-control without net magnetization~\cite{bai2024altermagnetism,fender2025altermagnetism,jungwirth2025altermagnetism,song2025altermagnets,jungwirth2026symmetry}. This novel phase has recently been realized in the van der Waals CsV$_2$Te$_2$O family, identified as a $d$-wave altermagnet in the 2D limit~\cite{zhang2025crystal,jiang2025metallic,hu2025pronounced,chen2025compression,Liu2025prbPhysical,Sun2025antiferrom,yang2025observation,wang2025atomic,fu2025atomic,jiaolin2026,hu2026observation}. In these systems, the Lieb lattice formed by the V$_2$O planes gives rise to quasi‑1D spin‑polarized flat bands and spin-valley locking Fermi surfaces~\cite{ma2021multifunctional}, which can in turn give rise to unconventional electromagnetic responses and spin‑dependent transport phenomena~\cite{li2025spinArxiv,Lai2025dwave,Zhang2025strain,yan2025sdw,Nagae25PRB}.

Despite its layered, device-friendly structure, the CsV$_2$Te$_2$O family poses a fundamental constraint for spin‑dependent applications: it is a G-type antiferromagnet in the bulk~\cite{Sun2025antiferrom,yang2025observation}, which preserves the combined inversion-time‑reversal symmetry and therefore guarantees spin-degenerate bulk bands. Nonetheless, the weak inter-layer coupling allows each layer to retain its local altermagnetic character~\cite{yang2025observation}. This gives rise to layer‑resolved altermagnetic spin‑split bands—a phase recently termed hidden altermagnetism or anti-altermagnetism—which can still host novel spin‑dependent phenomena~\cite{Matsuda2025multi,meier2025antiArxiv,guo2026hidden}. It remains an open question whether hidden altermagnet-based junctions can host superconducting phenomena with no analogue in other magnetic systems.

In this work, we demonstrate that the hidden altermagnetism in the CsV$_2$Te$_2$O family~\cite{yang2025observation} leads to pronounced altermagnetic even‑odd effects in both planar and vertical Josephson junctions with Rashba superconductors. In planar junctions with odd‑layer systems, the quasi‑1D spin‑polarized bands couple selectively to superconducting leads, producing a spin‑selective Josephson effect in which the critical supercurrent is fully spin‑polarized and exhibits strong directional anisotropy. In contrast, even‑layer structures exhibit complete cancellation of spin‑polarized supercurrents between top and bottom layers, suppressing this spin‑selectivity. In vertical junctions, odd-layer barriers promote equal-spin triplet transport and suppress the opposite-spin contribution, while even-layer barriers do the converse. Accordingly, the supercurrent alternates between larger and smaller values every added layer, producing a period-two modulation atop an overall decay with increasing barrier thickness. Our results establish hidden altermagnets as a unique platform for achieving gate‑tunable, spin‑polarized supercurrents without net magnetization, opening a new route toward superconducting spintronic devices.

\begin{figure}[t]
\centering
\includegraphics[width=0.98\linewidth]{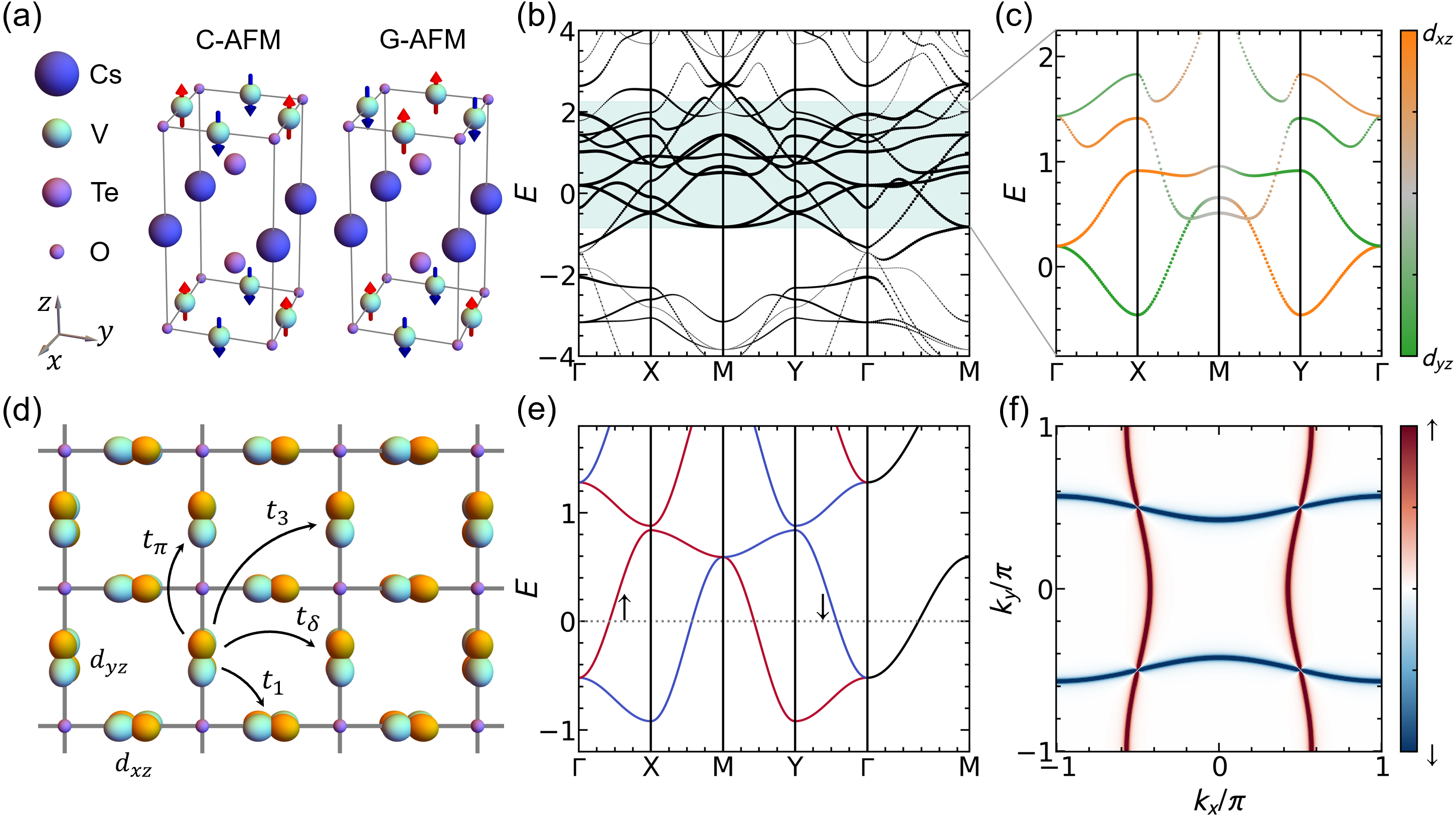}
\caption{Crystal and electronic structure of the $d$-wave altermagnetic CsV$_2$Te$_2$O.
(a) Magnetic lattice structure showing C-type or G-type antiferromagnetic orders.
(b) Nonmagnetic band structure of the monolayer; states near the Fermi level arise from V $3d$ orbitals (light teal region).
(c) Four-band projection near the Fermi level highlighting $d_{xz}$ and $d_{yz}$ orbitals.
(d) The effective V$_2$O Lieb lattice model with orbital-sublattice locking. The hopping parameters are labeled with black arrows.
(e) $d$-wave altermagnetic spin splitting bands from Eq.~\eqref{eq:H_AM}, with spins denoted by $\uparrow$ and $\downarrow$.
(f) Spin-resolved Fermi surface at the Fermi level [dashed line in (e)] featuring quasi-1D flat segments near $k_x, k_y \approx \pm \pi/2$.
}
\label{fig1}
\end{figure}

\vspace{0.5\baselineskip}
\noindent{\bf Material and Effective Model}\\
We focus on CsV$_2$Te$_2$O as our model system. Substitutions such as Rb or K for Cs do not introduce charge doping, and will not alter our main conclusions. The crystal structure belongs to the space group $P4/mmm$ (point group $D_{4h}$), with generators given by the four‑fold rotation $C_{4z}$, the two-fold rotation $C_{2x}$, and the out-of-plane reflection $M_z$. As illustrated in Fig.~\ref{fig1}(a), the V$_2$O planes form a Lieb lattice and support two competing magnetic orders. In both, the two V sublattices exhibit opposite spin polarization within the layer, an arrangement that can be interchanged only through rotation or reflection—the defining characteristic of altermagnetism~\cite{ma2021multifunctional}. The orders differ in their inter-layer coupling: the C‑type phase is ferromagnetic between layers [left panel in Fig.~\ref{fig1}(a)], whereas the G‑type phase is anti-ferromagnetically stacked [right panel in Fig.~\ref{fig1}(a)]. By symmetry, the C‑type phase is a bulk altermagnet, while the G‑type phase—in which neighboring layers host opposite altermagnetic spin‑splitting—corresponds to a hidden altermagnet~\cite{yang2025observation}. 

CsV$_2$Te$_2$O is a quasi-2D material with weak inter-layer coupling. We calculate its band structure using a monolayer model. The nonmagnetic band structure without spin-orbit coupling is shown in Fig.~\ref{fig1}(b). The electronic states near the Fermi level are dominated by the $3d$ orbitals of V atoms and are primarily shaped by the crystal field from the surrounding oxygen atoms. As revealed by quasi-particle interference measurements from scanning tunneling microscopy~\cite{wang2025atomic,fu2025atomic,jiaolin2026}, the low‑energy electronic structure is characterized by quasi‑1D flat bands. Our orbital‑resolved calculations in Fig.~\ref{fig1}(c), plotted along the high‑symmetry line, demonstrate that these experimentally observed bands derive primarily from the $d_{xz}$ and $d_{yz}$ orbitals~\footnote{The states at the $\Gamma$ point are two‑fold degenerate, corresponding to total angular momentum $\pm 1$: $\psi_{\Gamma,\pm}=\tfrac{1}{\sqrt{2}}(1,0,0,\pm i)$ or $\psi_{\Gamma,\pm}'=\tfrac{1}{\sqrt{2}}(0,1,\pm i,0)$. The time-reversal symmetry enforces this degeneracy. In contrast, the states at the $M$ point become non‑degenerate—even though they are also superpositions of $d_{xz}$ and $d_{yz}$ orbitals with equal weight—because they correspond to total angular momentum $0$ or $2$ (mod 4): $\psi_{M,0}=\tfrac{1}{2}(1,\pm1)\oplus(1,\pm1)$ or $\psi_{M,2}=\tfrac{1}{2}(1,\pm1)\oplus(1,\mp1)$.}.
Although the dispersion is not perfectly flat, the Fermi momentum lies near $\pi/2$, where $\cos k_x \approx 0$ or $\cos k_y \approx 0$, resulting in quasi‑1D Fermi surfaces with flat segments. Guided by this experimental and theoretical fingerprint, we construct the effective model based on these two dominant orbitals~\cite{cheng2026real}.

The crystal field further simplifies the model. Originating from the two oxygen atoms bonded to each V site along the bond direction, the local field significantly lifts the orbital degeneracy~\cite{jiang2025metallic}. This results in an on-site energy splitting $\Delta_{\text{CF}}=1.24$ eV between the $d_{xz}$ and $d_{yz}$ orbitals [Fig.~\ref{fig1}(c)]. Crucially, the sign of $\Delta_{\text{CF}}$ changes between the two symmetry-inequivalent V atoms within the unit cell. Due to its sizable magnitude, this crystal‑field configuration reduces the full Hilbert space [$(A,B)$ sublattices $\otimes$ $(d_{xz}$, $d_{yz})$ orbitals] to the low‑energy subspace spanned by the states $\vert A,d_{xz}\rangle$ and $\vert B, d_{yz}\rangle$ [Fig.~\ref{fig1}(d)]. Because of this sublattice-orbital locking, the sublattice index becomes redundant and will be omitted below. For the monolayer CsV$_2$Te$_2$O, the effective Hamiltonian in this two‑band subspace reads,
\begin{align} \label{eq:H_AM}
\begin{split}
{\cal H}_{\text{AM}}(\bm{k}) = [ & \epsilon_{2+}(\bm{k}) + \epsilon_3(\bm{k}) - \mu] \hat{\sigma}_0 \hat{s}_0 \\
+ & \epsilon_{2-}(\bm{k}) \hat{\sigma}_z \hat{s}_0 + \epsilon_1(\bm{k}) \hat{\sigma}_x \hat{s}_0 + M \hat{\sigma}_z \hat{s}_z,    
\end{split}
\end{align}
where the basis is $(d_{xz\uparrow}, d_{xz\downarrow}, d_{yz\uparrow}, d_{yz\downarrow})^T$ and $\hat{\sigma}$ ($\hat{s}$) are Pauli matrices for orbital (spin) degrees of freedom. $\bm{k} = (k_x,k_y)$ denotes the momentum in the plane. The coefficients are $\epsilon_{2\pm} = (t_\pi \pm t_\delta)(\cos{k_x} \pm \cos{k_y})$, $\epsilon_1 = -4t_1 \sin{\tfrac{k_x}{2}} \sin{\tfrac{k_y}{2}}$, and $\epsilon_3 = 4t_3 \cos{k_x} \cos{k_y}$. Here, $t_{\pi}$ and $t_{\delta}$ denote the intra‑sublattice hopping amplitudes from $dd\pi$ and $dd\delta$ bonds, respectively, $t_1$ is the nearest neighbor inter‑sublattice hopping, and $t_3$ is a next‑neighbor intra‑sublattice hopping [all illustrated in Fig.~\ref{fig1}(d)]. In addition, $M$ and $\mu$ denote the strength of the altermagnetic spin-splitting and the chemical potential, respectively. For $M<0$, the magnetic splitting shifts the $d_{xz,\downarrow}$ and $d_{yz,\uparrow}$ orbitals away from the Fermi level, as shown in Fig.~\ref{fig1}(e). This locking between atomic orbital and spin stems from the interplay between $\Delta_\text{CF}$ and $M$~\cite{wang2025prl,vila2024orbital}. Moreover, $M$ is on the order of $1$ eV, allowing us to neglect the spin-orbit coupling.

In this work, we adopt the parameter set (in units of eV): $t_\pi=-0.36$, $t_\delta=0.08$, $t_1=0.03$, $t_3=0.01$, $M=-0.9$, $\mu=-0.9$, which can reproduce the experimentally observed quasi-particle interference patterns~\cite{jiaolin2026}. The monolayer system exhibits a $d$-wave altermagnetic phase [Figs.~\ref{fig1}(e) and (f)], characterized by the spin‑space symmetry generators: $[U_s|C_{4z}]\triangleq i\hat{\sigma}_y\hat{s}_x$, $[U_s|C_{2[110]}]\triangleq \hat{\sigma}_x\hat{s}_x$. The altermagnetic nodal lines lie along the $[110]$ and $[1\bar{1}0]$ directions. In multilayer structures with G-type antiferromagnetic order~\cite{Sun2025antiferrom,yang2025observation}, the additional symmetry $[U_s|M_z]\triangleq\hat{\sigma}_0\hat{s}_x$ couples adjacent layers and enforces opposite altermagnetic spin‑splitting between $M_z$-related layers, defining the hidden altermagnetic phase~\cite{Matsuda2025multi,meier2025antiArxiv,guo2026hidden}. Consequently, $[U_s|M_z]$ is naturally broken in thin films with an odd number of layers, providing a clear symmetry distinction between even‑ and odd‑layer systems.

\begin{figure}[t]
\centering
\includegraphics[width=0.98\linewidth]{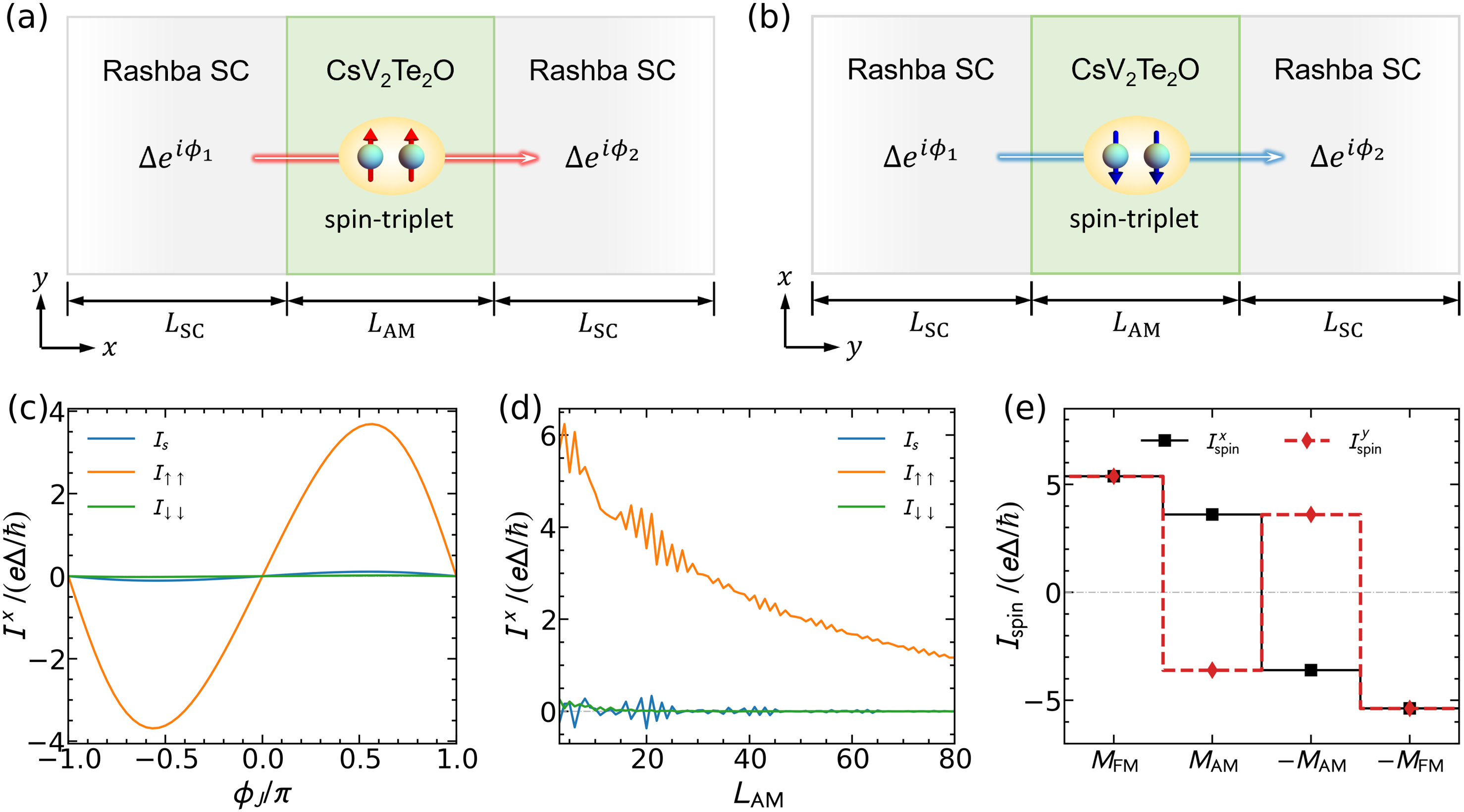}
\caption{Planar junctions in monolayer CsV$_2$Te$_2$O.
(a) and (b) illustrate the spin-selective Josephson effect in $s$-wave superconductor/altermagnet/$s$-wave superconductor Josephson junctions. It is along the (a) $x$ and (b) $y$ directions. The proximity-induced pairing correlations in the altermagnet are anisotropic: (a) purely spin-up triplet; and (b) purely spin-down triplet.
(c) For a short $x$-oriented junction ($L_\text{AM}=25$), the spin-singlet supercurrent $I^x_s$ is significantly smaller than the spin-triplet supercurrent $I^x_{\uparrow\uparrow}$, while $I^x_{\downarrow\downarrow}$ is vanishingly small.
(d) Dependence of $I^x_{s}$, $I^x_{\uparrow\uparrow}$, and $I^x_{\downarrow\downarrow}$ on the junction length $L_\text{AM}$.
(e) Comparison of the spin supercurrent for four different magnetic configurations: ferromagnetic order with $\pm M_\textbf{FM}\hat{\sigma}_0\hat{s}_z$ and altermagnetic order with $\pm M_\textbf{AM}\hat{\sigma}_z\hat{s}_z$. 
}
\label{fig2}
\end{figure}

\vspace{0.5\baselineskip}
\noindent{\bf Spin-selective Josephson effect}\\
We begin by considering the monolayer CsV$_2$Te$_2$O, and investigate its superconducting proximity effect when coupled to a conventional $s$-wave superconductor with Rashba spin-orbit coupling. Such Rashba superconductors are widely observed in non-centrosymmetric materials~\cite{smidman2017superconductivity}. The altermagnetic Fermi surface of CsV$_2$Te$_2$O consists of spin-polarized, nearly flat segments [Fig.~\ref{fig1}(f)]. Consequently, transport along a principal crystal axis is facilitated by only one spin species~\cite{Lai2025dwave}. In this work, we focus on Josephson junctions based on CsV$_2$Te$_2$O. When the junction is oriented along the $x$ direction [Fig.~\ref{fig2}(a)], the supercurrent is carried exclusively by spin-$\uparrow$ electrons, yielding a fully spin-polarized supercurrent. Strikingly, a simple $90^\circ$ rotation of the junction to the $y$ axis [Fig.~\ref{fig2}(b)] completely reverses this polarization. We term this directional, fully polarized superconducting transport the \textit{spin-selective Josephson effect}, a direct signature of the material's $d$-wave altermagnetic order.

To quantify the spin-selective Josephson effect, we employ the model in Eq.~\eqref{eq:H_AM} to compute the supercurrent. For Rashba superconductor/altermagnet/Rashba superconductor junctions, we model the $s$-wave superconducting leads with $C_{4z}$ symmetry,
\begin{align}
\begin{split}
{\cal H}_{\text{SC}}(\bm{k}) =& \epsilon_s(\bm{k}) \hat{\tau}_z\hat{s}_0 + \alpha(\sin k_x  \ \hat{\tau}_z\hat{s}_y - \sin k_y \ \hat{\tau}_0\hat{s}_x) \\
- &\Delta (\cos\phi \ \hat{\tau}_y\hat{s}_y +\sin\phi \ \hat{\tau}_x\hat{s}_y),
\end{split}
\end{align}
where the Nambu basis is $(c_{\bm{k},\uparrow},c_{\bm{k},\downarrow},c_{-\bm{k},\uparrow}^\dagger,c_{-\bm{k},\downarrow}^\dagger)^T$, and $\epsilon_s(\bm{k}) = -2t_s(\cos k_x + \cos k_y) -\mu_s$. Here, $t_s$, $\mu_s$, $\alpha$ are the hopping amplitude, chemical potential, and Rashba spin-orbit coupling strength, respectively; $\Delta$ and $\phi$ denote the $s$-wave pairing amplitude and the superconducting phase. We use parameters (in eV) $t_s=0.5$, $\alpha=0.4$, $\mu_s=-0.5$ and $\Delta=0.02$. This yields a superconducting coherence length of $\xi_\text{SC}\approx 15$. We confirm that our central results are robust and do not depend on this specific parameter set. Taking the junction along the $x$-direction as an example, the full Hamiltonian reads
\begin{align} \label{eq-hamJJ-x}
H_\text{JJ}(x,k_y) = H_\text{SC1} + H_\text{AM} + H_\text{SC2}+H_\text{coup},
\end{align}
where $H_\text{SC1}$ and $H_\text{SC2}$ describe the left and right superconducting leads (length $L_\text{SC}$) with phases $\phi_{1}$ and $\phi_{2}$, respectively, and $H_\text{AM}$ is the central altermagnet of length $L_\text{AM}$. The term $H_\text{coup}$ represents spin-independent hopping between the superconductors and the altermagnet. We assume translational invariance along $y$, while treating the $x$-direction with open boundaries and a finite number of layers. The explicit form of Eq.~\eqref{eq-hamJJ-x} is provided in the Methods section. Conservation of current allows the Josephson current to be calculated directly across the altermagnetic region~\cite{Asano2001PRB,Sakurai17PRB,SBZhang20PRB},
\begin{align} \label{eq-junction-Ic}
{I}_\text{tot}^x(\phi_J) =& -\frac{4e}{\hbar\beta} \sum_{k_y,\omega} \text{Im} \big[ \text{Tr} [ \hat{T}_h^\dagger \hat{F}(x+1) \hat{T}_e \hat{F}'(x)] \big],
\end{align}
where $\phi_J=\phi_1-\phi_2$ is the phase difference, $\beta = 1/k_B T$, $\omega=(2n+1)\pi/\beta$ are the fermionic Matsubara frequencies. $\hat{T}_{e/h}(k_y)$ denote the electron (hole) hopping matrices, while $\hat{F}(x,k_y,\omega)$ and  $\hat{F}'(x,k_y,\omega)$ represent the bulk and surface anomalous Green's functions at site $x$, respectively. These two matrices encode the proximity‑induced pair correlations within the altermagnet, directly linking them to the supercurrent via Eq.~\eqref{eq-junction-Ic}. In the spin basis, $\hat{F}(x,k_y,\omega)$ decomposes into singlet and triplet components: $\hat{F} = -i\hat{F}_s\hat{s}_y  +\hat{F}_{\uparrow\uparrow}\tfrac{\hat{s}_0+\hat{s}_z}{2}  +\hat{F}_{\downarrow\downarrow}\tfrac{\hat{s}_0-\hat{s}_z}{2}  +\hat{F}_z\hat{s}_x$, where $\hat{F}_s$, $\hat{F}_{\uparrow\uparrow}$, $\hat{F}_{\downarrow\downarrow}$, and $\hat{F}_z$ are $2\times2$ matrices in the orbital subspace, corresponding to spin-singlet, equal-spin-triplet, and opposite-spin-triplet correlations, respectively. Crucially, we find that only the $\hat{F}_{\uparrow\uparrow}$ channel develops a non-zero proximity-induced pairing within the altermagnet, manifesting a fully spin-selective superconducting effect (see Methods). Consequently, the total Josephson current in Eq.~\eqref{eq-junction-Ic} naturally separates into singlet, triplet, and mixed contributions, 
\begin{align}
I_{\text{tot}}^{x}(\phi_J) = I_{s}^x + I_{z}^x + I_{\uparrow\uparrow}^x + I_{\downarrow\downarrow}^x +  I_{\text{mix}}^x.
\end{align}
Due to the spin U(1) symmetry of the altermagnet, the mixed term $I_{\text{mix}}^x$ arises solely as a cross term between the spin-singlet and opposite-spin-triplet correlations. And it can only be activated via spin-orbit coupling within the altermagnet. The supercurrent for the $y$-oriented junction, $I_{\text{tot}}^{y}(\phi_J)$, is obtained analogously.

Figure~\ref{fig2}(c) shows the current‑phase relations for each component in a short $x$‑direction junction ($L_\text{AM}=25$). Both the spin-singlet supercurrent $I_s^x$ (blue curve) and the spin-$\uparrow$ polarized triplet supercurrent $I_{\uparrow\uparrow}^x$ (orange curve) are present. As expected, $I_s^x$ is significantly smaller than $I_{\uparrow\uparrow}^x$. Strikingly, despite the zero net magnetization of the altermagnet, $I_{\uparrow\uparrow}^x$ dominates the transport, while $I_{\downarrow\downarrow}^x$ (green curve) remains vanishingly small. This dominance originates from the nearly flat spin‑$\downarrow$ Fermi surface segment, whose vanishingly small Fermi velocity strongly suppresses its contribution to charge transport, leaving only the spin-$\uparrow$ band to carry the supercurrent. Moreover, the opposite-spin triplet term $I_z^x$ is comparable in magnitude to $I_s^x$, while the mixing term $I_\text{mix}^x$ vanishes identically (see Methods).

At a fixed $\phi_J=\pi/2$, we plot the supercurrent $I^x_{s/\uparrow\uparrow/\downarrow\downarrow}$ versus junction length $L_\text{AM}$ [Fig.~\ref{fig2}(d)]. Both $I_s^x$ and $I_{\downarrow\downarrow}^x$ stay small and decay rapidly with distance, whereas the spin-$\uparrow$ channel persists. Hence, the Josephson supercurrent is fully spin‑polarized over a wide range of junction lengths. For the $y$-oriented junction, the spin‑space symmetry $[U_s|C_{4z}]$ yields analogous results, but with only the spin-$\downarrow$ channel carrying the supercurrent. Thus, for CsV$_2$Te$_2$O-based junctions, we conclude:
\begin{align} \label{eq-ssje}
\begin{cases}
x\text{-direction junction, only } I_{\uparrow\uparrow}^x \text{ is finite}, \\
y\text{-direction junction, only } I_{\downarrow\downarrow}^y \text{ is finite}.
\end{cases}    
\end{align}
We term this phenomenon the \textit{spin-selective Josephson effect}. Moreover, $[U_s|C_{4z}]$ enforces $I_{\uparrow\uparrow}^x = I_{\downarrow\downarrow}^y$. These results establish a route toward dissipationless spin control in superconducting spintronic devices.

To quantify the spin‑selectivity and distinguish the altermagnetic state from ferromagnetic orders, we define the spin supercurrents for the two junction orientations,
\begin{align}
I_{\text{spin}}^{x/y}=I_{\uparrow\uparrow}^{x/y} -I_{\downarrow\downarrow}^{x/y}, 
\end{align}
which directly probes the altermagnetic response to an external magnetic field. For clarity, we consider four representative magnetic configurations (neglecting transition processes): (i) a ferromagnet with majority spin $\uparrow$; (ii) an altermagnet with N\'eel vector along $+\hat{z}$; (iii) an altermagnet with N\'eel vector along $-\hat{z}$; and (iv) a ferromagnet with majority spin $\downarrow$. These phases can, in principle, be tuned by an out‑of‑plane magnetic field applied to the CsV$_2$Te$_2$O layer, assuming the superconducting leads remain essentially unaffected~\footnote{To our knowledge, the critical field for the altermagnet-to-ferromagnet transition in CsV$_2$Te$_2$O has not yet been experimentally determined.}.
The results are summarized in Fig~\ref{fig2}(e). In the altermagnetic state, the spin supercurrents along the two orthogonal directions are opposite in sign, $I_{\text{spin}}^x =- I_{\text{spin}}^y$. Reversing the N\'eel vector flips the sign of both $I_{\text{spin}}^x$ and $I_{\text{spin}}^y$. In contrast, when the field drives the system into a ferromagnetic state, the two spin supercurrents become equal, indicating that the spin‑polarized supercurrent propagates isotropically, independent of the junction orientation.

Therefore, we conclude that the spin-selective Josephson effect constitutes a unique signature of the CsV$_2$Te$_2$O family hidden altermagnets, distinct from all previously reported behaviors in altermagnet‑based Josephson junctions~\cite{Ouassou2023prl,Zhang2024NatComm,Beenakker2023prb,Cheng2024prb,SunHP25PRB,LuB24PRL,Sharna25PRB,Amartya25PRB,ZhaoWJ25PRB,li2025spinArxiv,YangZX2502,Heinsdorf2026prb}.

\begin{figure}[t]
\centering
\includegraphics[width=0.98\linewidth]{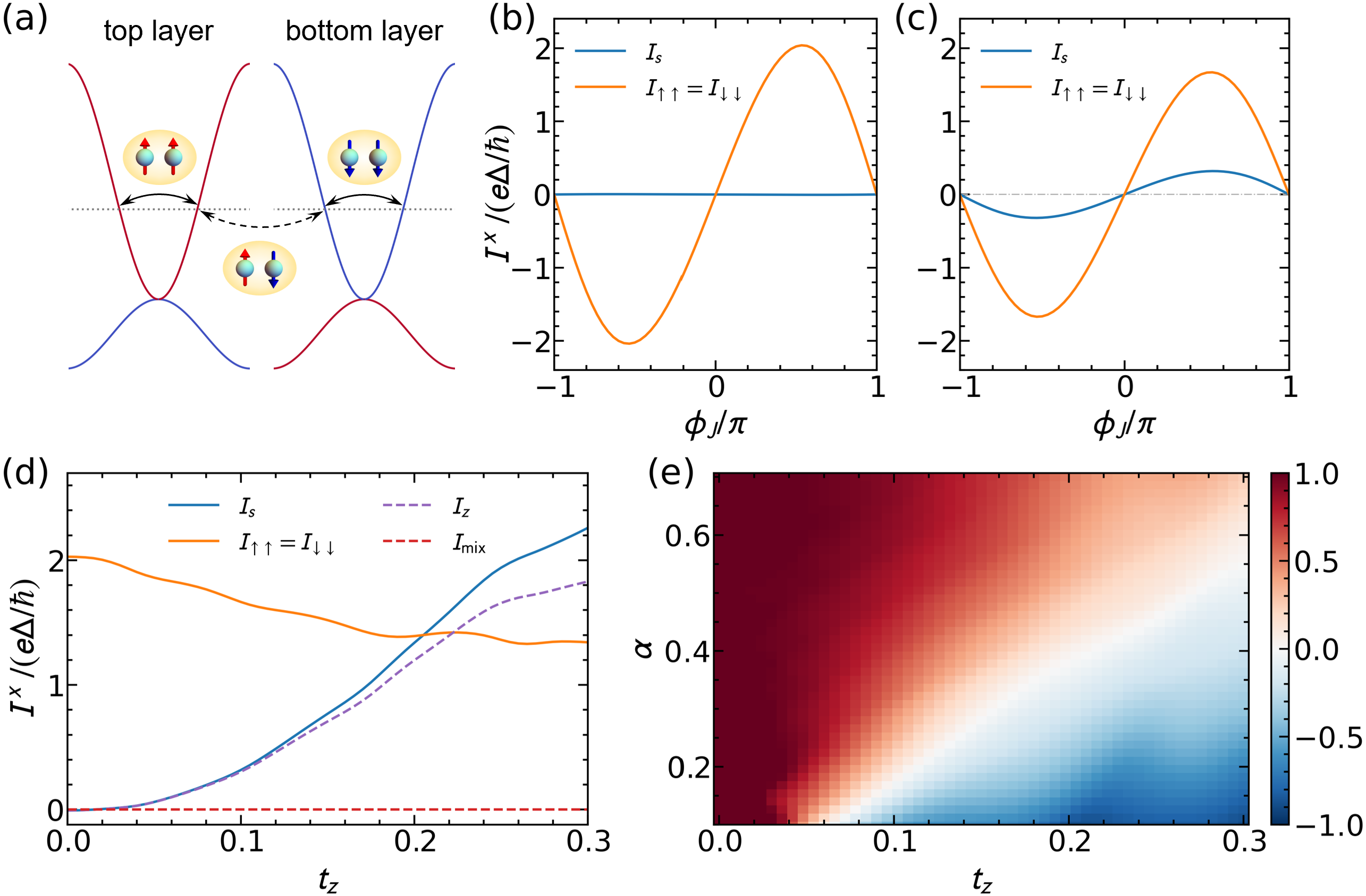}
\caption{Planar junctions in bilayer CsV$_2$Te$_2$O.
(a) Schematic of layer-resolved spin-split bands in the top and bottom layers. Proximity-induced superconducting pairings include intra-layer equal-spin triplet and inter-layer opposite-spin triplet pairing, the latter mediated by the inter-layer hopping $t_z$.
(b)-(c) Supercurrent components in a long junction ($L_\text{AM}=50$) for two cases:
(b) For $t_z=0.01$ eV, the spin-singlet supercurrent $I_s^x$ is negligible.
(c) For $t_z=0.1$ eV, $I_s^x$ becomes comparable to the equal-spin triplet components $I_{\uparrow\uparrow}^x=I_{\downarrow\downarrow}^x$.
(d) Dependence of all supercurrent components on $t_z$ at $\alpha=0.4$ and $\phi_J = \pi/2$.
(e) Normalized supercurrent difference $\Delta I^x_\text{JJ}/I^x_\text{tot}$ in the $t_z$-$\alpha$ plane, where $\Delta I^x_\text{JJ}$ is defined in Eq.~\eqref{eq-Ix-JJ-diff}.
}
\label{fig3}
\end{figure}

\vspace{0.5\baselineskip}
\noindent{\bf Phase diagram for bilayer system}\\
We next study the bilayer system. As noted, the symmetry $[U_s|M_z]$ distinguishes the monolayer from the bilayer system—the latter being another important structural unit for CsV$_2$Te$_2$O. This symmetry ``hides'' the altermagnetic spin-splitting in the bilayer, thereby suppressing the \textit{spin-selective Josephson effect} found in the monolayer [Eq.~\eqref{eq-ssje}]. Nevertheless, an equal-spin-triplet supercurrent persists for bilayer junctions along both the $x$ and $y$ directions. As illustrated in Fig.~\ref{fig3}(a), equal-spin-triplet pairing correlations are induced separately in each layer: $\hat{F}_{\uparrow\uparrow}$ in the top layer and $\hat{F}_{\downarrow\downarrow}$ in the bottom layer. Crucially, additional supercurrent-carrying channels arise from inter-layer spin-singlet and opposite-spin-triplet pairing correlations. These channels are controlled by the inter-layer hopping $t_z$. For $t_z=0$, the bilayer behaves as two independent altermagnetic layers and the additional channels vanish, as expected. As $t_z$ increases to values comparable to the intra-layer hoppings, the system crosses over to the bulk limit, where singlet and triplet pairings coexist. To investigate this $t_z$-driven crossover, we consider the bilayer Hamiltonian
\begin{align}
\begin{split}
{\cal H}_{\text{bAM}}(\bm{k}) = [ &\epsilon_{2+}+ \epsilon_3 - \mu] \hat{\gamma}_0 \hat{\sigma}_0 \hat{s}_0 + \epsilon_{2-} \hat{\gamma}_0 \hat{\sigma}_z \hat{s}_0 \\
+ & \epsilon_1 \hat{\gamma}_0 \hat{\sigma}_x \hat{s}_0 + M \hat{\gamma}_z \hat{\sigma}_z \hat{s}_z + t_z\hat{\gamma}_x\hat{\sigma}_0\hat{s}_0,
\end{split}
\end{align}
where $\hat{\gamma}_\mu$ are Pauli matrices acting on the layer degree of freedom. The energy spectrum remains spin-degenerate. We calculate the Josephson supercurrent along the $x$ direction for the bilayer junction. To suppress intra-layer contributions to the spin-singlet supercurrent, we consider a long junction of length $L_\text{AM}=50$. In the weak inter-layer coupling regime ($t_z=0.01$ eV), the supercurrent is dominated by equal-spin-triplet pairings, with a negligible spin-singlet contribution $I_s^x$ [Fig.~\ref{fig3}(b)]. Notably, we confirm $I_{\uparrow\uparrow}^x=I_{\downarrow\downarrow}^x$, a direct consequence of the G-type antiferromagnetic order. When $t_z$ increases to $0.1$ eV, $I_s^x$ becomes significant, though still smaller than $I_{\uparrow\uparrow}^x$ [Fig.~\ref{fig3}(c)]. This indicates that enhanced inter-layer hybridization can dramatically tune the supercurrent composition in bilayer junctions.

To elucidate the microscopic origin of this evolution, Fig.~\ref{fig3}(d) shows all supercurrent components $I^x_{m}(\pi/2)$ with $m=\{s,\uparrow\uparrow,\downarrow\downarrow,z,\text{mix}\}$ as a function of $t_z$. As $t_z$ increases, the intra-layer equal-spin-triplet supercurrents ($I_{\uparrow\uparrow}^x=I_{\downarrow\downarrow}^x$) remain robust, while both the spin-singlet supercurrent $I_s^x$ and the opposite-spin-triplet supercurrent $I_z^x$ are progressively enhanced. Moreover, since $t_z\le 0.3$ eV remains smaller than the altermagnetic spin-splitting strength ($|M|=0.9$ eV), the equal-spin-triplet supercurrents exhibit only weak variation. However, $I_\text{mix}^x$ remains exactly zero owing to the absence of spin-dependent hopping parameters in the altermagnet (see Methods).

Beyond the tuning of singlet-triplet components via $t_z$, the Rashba spin-orbit coupling $\alpha$ in the $s$-wave superconducting leads also governs the conversion between singlet and triplet pairings. To map out the supercurrent composition, we construct a two-dimensional phase diagram in the $t_z$-$\alpha$ plane [Fig.~\ref{fig3}(e)]. We quantify the competition by defining
\begin{align} \label{eq-Ix-JJ-diff}
\Delta I^x_{\text{JJ}} = I^x_\text{eq} - I^x_\text{op},
\end{align}
where $I^x_\text{eq} = I^x_{\uparrow\uparrow} + I^x_{\downarrow\downarrow}$ and $I^x_{\text{op}}=I^x_s+I^x_z+I^x_\text{mix}$ denote the supercurrent contributions from pairings with total spin angular momentum $\pm 1$ and $0$, respectively. 
Figure~\ref{fig3}(e) shows the normalized supercurrent difference $\Delta I^x_{\text{JJ}}/I^x_\text{tot}$. The Josephson supercurrent is dominated by equal-spin-triplet pairings in the regime of large $\alpha$ but small $t_z$ ($I^x_\text{eq} \gg I^x_\text{op}$). In contrast, for small $\alpha$ but large $t_z$, the spin-singlet supercurrent becomes dominant ($I^x_\text{op}\approx I^x_s \gg I^x_\text{eq}$). In practice, uniaxial strain along the $[001]$ direction could provide a way to control $t_z$, enabling a controlled crossover between triplet-dominated and singlet-dominated supercurrents.

The inter-layer hopping in unstrained CsV$_2$Te$_2$O is only $0.01$ eV~\cite{yang2025observation}, firmly placing the system within the weak-coupling regime. Therefore, we conclude that equal-spin-triplet supercurrents dominate in bilayer CsV$_2$Te$_2$O-based Josephson junctions, despite the material being in a G-type antiferromagnetic phase.

\begin{figure}[t]
\centering
\includegraphics[width=0.98\linewidth]{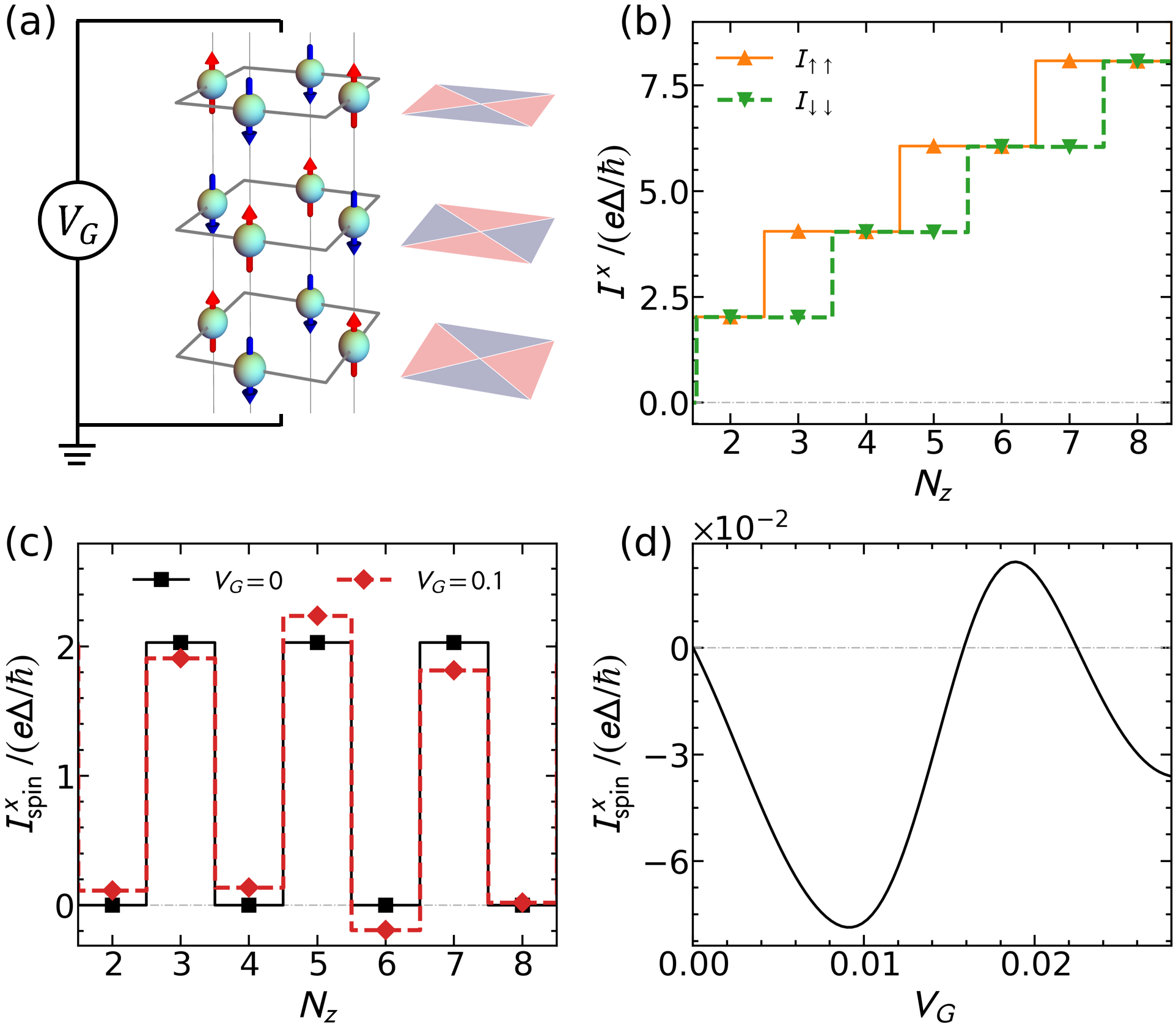}
\caption{Altermagnetic even-odd effect in CsV$_2$Te$_2$O with small inter-layer hopping $t_z=0.01$ eV.
(a) Schematic illustration of the planar junction for the multilayer structure. A gate voltage is applied perpendicular to the altermagnetic planes.
(b) Layer-parity dependence of $I^x_{\uparrow\uparrow}$ and $I^x_{\downarrow\downarrow}$ at $\phi_J=\pi/2$.
(c) The corresponding spin current $I^x_\text{spin} = I^x_{\uparrow\uparrow}-I^x_{\downarrow\downarrow}$, showing on-off switching by adding layers ($V_G=0$, black curve). A finite gate voltage ($V_G = 0.1$ eV) induces a spin supercurrent in even-$N_z$ systems (red curve).
(d) Nonmonotonic dependence of the spin current $I^x_\text{spin}$ on gate voltage $V_G$ for a four-layer junction ($N_z = 4$).
}
\label{fig4}
\end{figure}

\vspace{0.5\baselineskip}
\noindent{\bf Altermagnetic even-odd effect}\\
The preceding sections reveal a striking and, to our knowledge, previously unrecognized phenomenon: the Josephson effect in hidden altermagnets is governed fundamentally by layer parity. In the monolayer, the altermagnetic spin splitting yields a fully spin-selective supercurrent—only $I_{\uparrow\uparrow}^x$ (or $I_{\downarrow\downarrow}^y$) carries the charge. In the bilayer, this selectivity is hidden by the symmetry $[U_s|M_z]$, yet equal-spin-triplet supercurrents ($I_{\uparrow\uparrow}^x = I_{\downarrow\downarrow}^x$ and $I_{\uparrow\uparrow}^y = I_{\downarrow\downarrow}^y$) persist and dominate as long as the inter-layer hopping remains weak. We unify these distinct phenomena under the concept of an \textit{altermagnetic even‑odd effect}: the layer‑number parity fundamentally dictates the symmetry and spin polarization of the supercurrent in multilayer systems. Thus, we consider an $N_z$-layered CsV$_2$Te$_2$O films described by,
\begin{align} \label{eq-ham-multiL}
\begin{split}
{\cal H}_{\text{mAM}}(\bm{k}) = [ &\epsilon_{2+}+ \epsilon_3 - \mu] \hat{\Gamma}_0 \hat{\sigma}_0 \hat{s}_0 + \epsilon_{2-} \hat{\Gamma}_0 \hat{\sigma}_z \hat{s}_0 \\
+ & \epsilon_1 \hat{\Gamma}_0 \hat{\sigma}_x \hat{s}_0 + M \hat{\Gamma}_z \hat{\sigma}_z \hat{s}_z + t_z\hat{\Gamma}_x\hat{\sigma}_0\hat{s}_0,
\end{split}
\end{align}
where $\hat{\Gamma}_0$ is the $N_z\times N_z$ identity matrix, $\hat{\Gamma}_z=\text{diag}[1, -1, 1, -1, \dots, (-1)^{N_z+1}]$ encodes the alternating layer magnetization, and $\hat{\Gamma}_x$ is an $N_z \times N_z$ symmetric tridiagonal matrix representing inter-layer hopping (ones on the first sub‑ and super‑diagonals, zeros elsewhere). Notice that the system carries zero net magnetization. Using a multilayer Josephson junction platform with an electrostatic gate $V_G$ [Fig.~\ref{fig4}(a)], we demonstrate how the superconducting transport evolves with layer parity, thereby establishing the \textit{altermagnetic even‑odd effect} as a general and tunable design principle for altermagnet‑based superconducting devices.

We now focus on $x$-oriented junctions. In Fig.~\ref{fig4}(b), we present the equal-spin-triplet supercurrent components, $I_{\uparrow\uparrow}^x(N_z)$ and $I_{\downarrow\downarrow}^x(N_z)$ at $\phi_J=\pi/2$, as functions of $N_z$. For even $N_z$, the two components are exactly equal, $I_{\uparrow\uparrow}^x(N_z) = I_{\downarrow\downarrow}^x(N_z)$, yielding a vanishing net spin supercurrent, $I_\text{spin}^x(N_z)=0$. The spin‑singlet contributions remain negligible as we consider the small inter‑layer coupling $t_z$. For odd $N_z$, the spin balance is broken, and the supercurrent obeys the following parity relations,
\begin{subequations}
\begin{align}
I_{\uparrow\uparrow}^x(N_z) &= I_{\uparrow\uparrow}^x(N_z+1), \\
I_{\downarrow\downarrow}^x(N_z) &= I_{\downarrow\downarrow}^x(N_z-1).
\end{align}    
\end{subequations}
Likewise, junctions oriented along the $y$ direction exhibit fully analogous behavior, with the roles of spin‑up and spin‑down currents exchanged. Thus, the relative magnitude of $I_{\uparrow\uparrow}^{x/y}$ and $I_{\downarrow\downarrow}^{x/y}$ is a sensitive probe of layer parity—directly reflecting the alternating spin texture of the hidden altermagnetic order in CsV$_2$Te$_2$O films.

To further quantify the even‑odd switching, we plot the net spin supercurrent $I_\text{spin}^{x}$ in Fig.~\ref{fig4}(c). The layer number $N_z$ acts as a spin‑current switch (solid curve): $I_\text{spin}^{x}$ is finite for odd $N_z$ (``on'') and vanishes for even $N_z$ (``off''). This parity‑controlled signal can be further manipulated by an external gate voltage $V_G$~\cite{peng2025all}, as illustrated in Fig.~\ref{fig4}(a), which breaks the spin-space symmetry $[U_s|M_z]$ and thereby activates a finite $I_\text{spin}^{x}$ even in even-$N_z$ systems (see Methods for modeling details). The dashed curve in Fig.~\ref{fig4}(c) shows the result for $V_G=0.1$ eV. Strikingly, the gate-induced spin supercurrent in even-$N_z$ systems exhibits a nonmonotonic dependence on thickness. This behavior originates from a more fundamental nonmonotonic response of $I_\text{spin}^{x}$ to $V_G$ [see Fig.~\ref{fig4}(d) for the $N_z=4$ case], which arises from finite-momentum Cooper pairing.

The altermagnetic even‑odd effect in the CsV$_2$Te$_2$O family is distinct from the even‑odd effects reported in topological antiferromagnetic MnBi$_2$Te$_4$ films~\cite{chen2019intrinsic,zhang2019experimental,ovchinnikov2021intertwined,gao2021layer,Yang2021odd,zhao2021even}. The latter originates from surface-state topology and exhibits a net magnetization in odd layers that vanishes in even layers; the former, in contrast, arises purely from hidden altermagnetism and yields zero net magnetization in both odd and even layers. Namely, the even-odd effect in MnBi$_2$Te$_4$ is reflected in its net magnetization, while in CsV$_2$Te$_2$O it is reflected in its net altermagnetism. Our results thus establish an entirely different physical mechanism and a new material platform for parity-controlled superconducting spintronics.

\vspace{0.5\baselineskip}
\noindent{\bf Discussion and conclusion} 

\textit{Material realization and experimental signatures.--}
This CsV$_2$Te$_2$O family is believed to be hidden altermagnet, and our results on CsV$_2$Te$_2$O can be applied to other materials in this family. Our effective model for monolayer CsV$_2$Te$_2$O includes only the $d_{xz}$ and $d_{yz}$ orbitals and omits the Fermi surfaces near $X$ and $Y$ points. We discuss three points that support the robustness of our conclusions despite this simplification: 
(i) The impurity‑induced bound states recently observed by scanning tunneling microscopy~\cite{wang2025atomic,fu2025atomic,jiaolin2026} originate from quasi-1D near‑flat segments of the Fermi surface. 
(ii) These additional Fermi surfaces can, in principle, be eliminated via electrostatic gating or chemical substitution in altermagnetic thin films.
(iii) Even if present, these Fermi surfaces also carry spin‑polarized supercurrents that are fully consistent with—and do not alter—our central conclusions regarding the spin‑selective and even‑odd Josephson effects. More broadly, the altermagnetic even‑odd effect is not limited to Josephson junctions; it can be extended to a wide range of magnetic and transport phenomena, including spin injection, magnetoresistance, and nonlocal correlations. The even-odd effect in Josephson junctions may be directly accessible via differential conductance spectra~\cite{Hubler2012observation} and microwave spectroscopy~\cite{Tosi2019spin}, or indirectly probed through spin-polarized Andreev bound states detected by scanning tunneling microscopy~\cite{Wang2025quantum}—a direction we leave for future work.

\begin{figure}[b]
\centering
\includegraphics[width=0.98\linewidth]{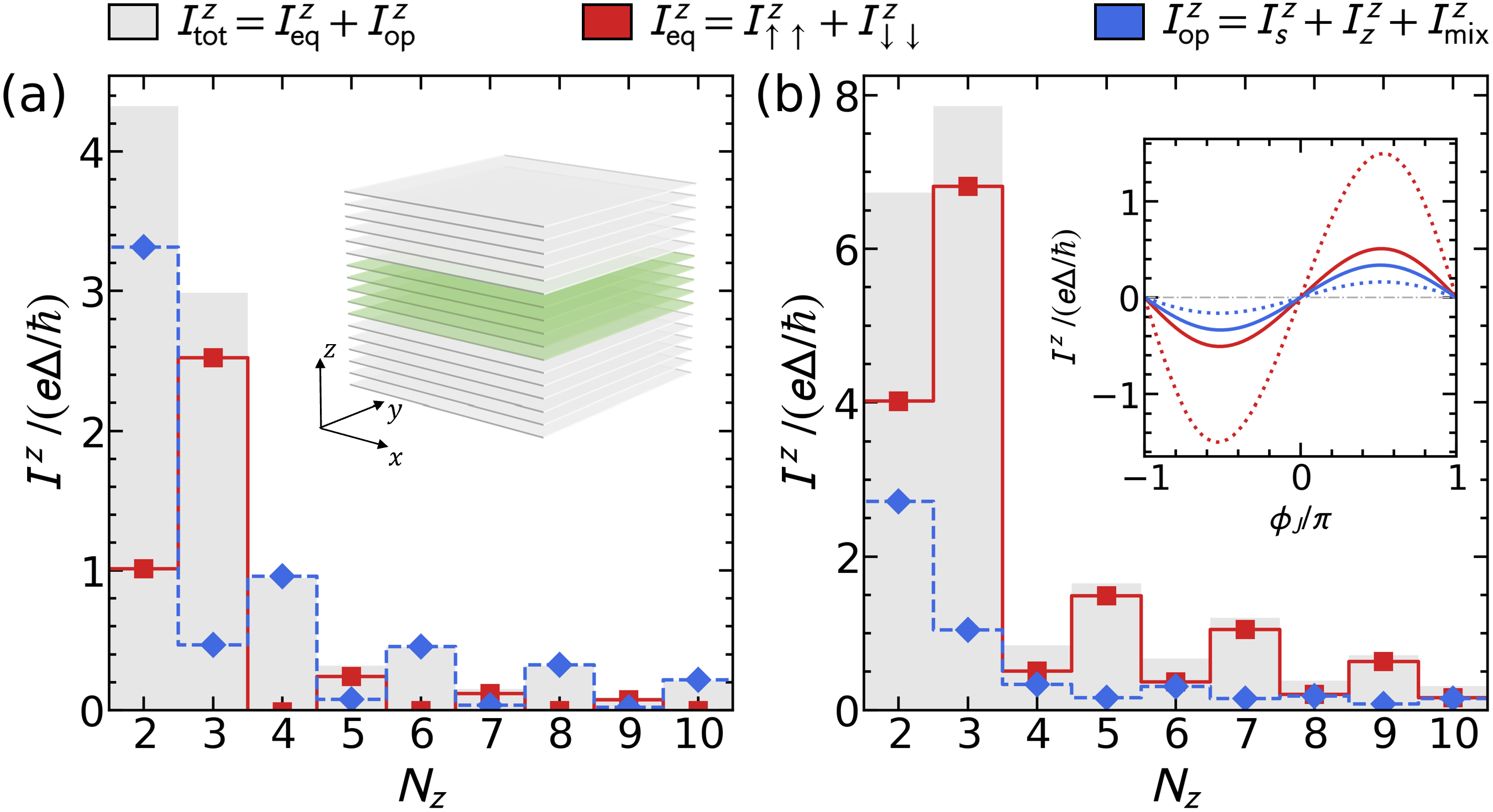}
\caption{Altermagnetic even-odd effect in vertical Josephson junctions.
(a) Layer dependence of the equal-spin triplet supercurrent  $I^z_{\text{eq}}$ and opposite-spin supercurrent $I^z_{\text{op}}$ at $\phi_J=\pi/2$ and $\alpha=0.1$. Gray-shaded rectangles represent the total supercurrent. Inset: schematic of the vertical junction.
(b) Same as (a) but for $\alpha=0.4$. Inset: current-phase relation for $N_z = 4$ (solid curve) and $N_z = 5$ (dotted curve). Here, $t_z = 0.2$ is used to emphasize the even-odd effect.
}
\label{fig5}
\end{figure}

\textit{Even-odd effect in vertical junctions.--} 
While the preceding analysis focuses on planar junctions, the altermagnetic even‑odd effect is not limited to this geometry. Vertical Josephson junctions—where the supercurrent flows perpendicular to the altermagnetic layers—offer an alternative and technologically relevant platform. When the inter-layer hopping $t_z$ is sufficiently large, transport along the vertical direction becomes appreciable. To be concrete, we consider a Josephson junction oriented along $z$, in which an $N_z$-layer altermagnet with G-type antiferromagnetic order [Eq.~\eqref{eq-ham-multiL}] is sandwiched between two Rashba superconductors [Fig.~\ref{fig5}(a), inset]. We assume in-plane translational symmetry so that $k_x$ and $k_y$ remain good quantum numbers. In this geometry, the symmetry $[U_s\Vert C_{4z}]$ is preserved. Consequently, the two spin channels are symmetry-related and no spin-selective Josephson effect appears. Nevertheless, the interplay between $t_z$ in the altermagnet and $\alpha$ in the Rashba superconducting leads enables both equal-spin triplet and opposite-spin pairing correlations to be induced and transmitted across the junction. This suggests the possibility of a pronounced even-odd layer dependence. To explore this, we calculate the equal-spin triplet ($I^z_\text{eq} = I^z_{\uparrow\uparrow} + I^z_{\downarrow\downarrow}$) and opposite-spin ($I^z_{\text{op}}=I^z_s+I^z_z+I^z_\text{mix}$) contributions to the supercurrent at $\phi_J=\pi/2$ as functions of $N_z$ [Fig.~\ref{fig5}].

Our calculations reveal a clear even-odd dichotomy:  odd-$N_z$ junctions enhance $I^z_\text{eq}$ while suppressing $I^z_{\text{op}}$. This behavior stems from the net altermagnetic spin splitting present in odd-layer barriers, which facilitates the propagation of equal-spin triplet pairs but disrupts opposite-spin transport. By contrast, even-$N_z$ junctions host spin-degenerate bands, and thus favor opposite-spin pairing. These even-odd contrasts are robust in the weak spin-orbit coupling regime: $\alpha<t_z$.

Explicitly, $I^z_\text{eq}$ is enhanced for odd $N_z$ but strongly suppressed for even $N_z$ [blue solid line in Fig.~\ref{fig5}(a)]. Conversely, $I^z_{\text{op}}$ is enhanced for even $N_z$ and suppressed for odd $N_z$ [red dashed line in Fig.~\ref{fig5}(a)]. As a result, on top of the overall decay, the total current $I_\text{tot}^z=I_\text{eq}^z+I_\text{op}^z$ oscillates with period two as increasing $N_z$: larger for even layers and smaller for odd layers.

This even-odd effect persists in the strong spin-orbit coupling regime ($\alpha>t_z$), with the key distinction that equal-spin triplet pairs now carry the majority of the supercurrent [Fig.~\ref{fig5}(b)]. Consequently, the total current becomes stronger for odd layers and weaker for even layers. Although we present results at $\phi_J=\pi/2$, the even-odd effect holds across other phase differences, as evidenced by the current-phase relations in the inset of Fig.~\ref{fig5}(b). For clarity, we present the results at an enhanced $t_z$; the same qualitative features persist for smaller $t_z$, albeit with faster decay as $N_z$ increases.

\textit{Conclusion.--}
We have uncovered a family of altermagnetic even-odd effects in Josephson junctions based on CsV$_2$Te$_2$O—a van der Waals $d$-wave altermagnet with G-type antiferromagnetic order. In planar junctions, the quasi‑1D, spin-polarized Fermi surfaces with flat segments yields a fully spin-polarized, directionally anisotropic supercurrent that survives only in odd-layer systems; even-layer junctions exhibit exact cancellation of the spin supercurrent. Thus, layer parity acts as a switch for spin-polarized supercurrent. In vertical junctions, odd layers favor equal-spin triplet transport while even layers favor opposite-spin transport, giving rise to a robust period-two oscillation in the total supercurrent. These layer-parity-controlled responses constitute a generic even-odd phenomenon intrinsic to hidden altermagnets, distinct from conventional even-odd effects in antiferromagnetic or nonmagnetic multilayers. Our theory can also be applied to other magnetic materials, such as X-type antiferromagnets~\cite{Zhang2025xtype}.

\begin{acknowledgments}
We thank G.-W.~Yang, X.~Lu, L.~Jiao, H.-M.~Yi, Z.-Y.~Zhang and F.-C.~Zhang for helpful discussions. 
C.L. and L.H.H. were supported by National Key R\&D Program of China (Grant No. 2025YFA1411501), the National Natural Science Foundation of China (Grant Nos. 12561160109, 1257040632), the Fundamental Research Funds for the Central Universities (Grant No. 226-2024-00068).
C.L. was also supported by central fiscal special-purpose fund (Grant No. 2021ZD0302500).
J.X.H. and S.B.Z. were supported by the start-up fund at HFNL, the National Natural Science Foundation of China (Grant No.~12488101) and the Innovation Program for Quantum Science and Technology (Grant No. 2021ZD0302801).
\end{acknowledgments}

\bibliography{refs2026}
\end{document}